\documentclass[Physsubmission, Phys]{SciPost}

\hypersetup{
    colorlinks,
    linkcolor={red!50!black},
    citecolor={blue!50!black},
    urlcolor={blue!80!black}
}

\usepackage[bitstream-charter]{mathdesign}
\DeclareSymbolFont{usualmathcal}{OMS}{cmsy}{m}{n}
\DeclareSymbolFontAlphabet{\mathcal}{usualmathcal}

\begin{document}

\begin{center}{\Large \textbf{Cosmic ray interactions in the atmosphere:\\
QGSJET-III and other models}}\end{center}

\begin{center}
S.\ Ostapchenko\textsuperscript{1,2$\star$}
\end{center}

\begin{center}
{\bf 1}  Universit\"at Hamburg, II Institut f\"ur Theoretische
Physik, 22761 Hamburg, Germany
\\
{\bf 2}  D.V. Skobeltsyn Institute of Nuclear Physics,
Moscow State University, 119992 Moscow, Russia
\\
* sergey.ostapchenko@desy.de
\end{center}

\begin{center}
\today
\end{center}


\definecolor{palegray}{gray}{0.95}
\begin{center}
\colorbox{palegray}{
  \begin{tabular}{rr}
  \begin{minipage}{0.1\textwidth}
    \includegraphics[width=30mm]{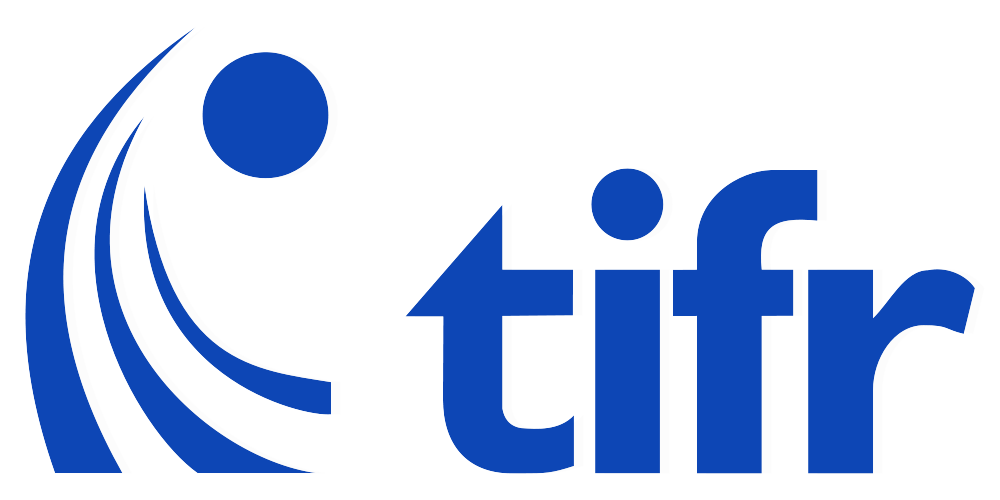}
  \end{minipage}
  &
  \begin{minipage}{0.85\textwidth}
    \begin{center}
    {\it 21st International Symposium on Very High Energy Cosmic Ray 
    Interactions (ISVHECRI 2022)}\\
    {\it Online, 23-27 May 2022} \\
    \doi{10.21468/SciPostPhysProc.?}\\
    \end{center}
  \end{minipage}
\end{tabular}
}
\end{center}

\section*{Abstract}
{\bf
The physics content of the QGSJET-III Monte Carlo generator of high energy
hadronic collisions is discussed. New theoretical approaches implemented in 
 QGSJET-III are addressed
in some detail and a comparison to alternative treatments
 of other cosmic ray interaction models is performed. Calculated
  characteristics of cosmic ray-induced extensive air showers are presented
  and differences between the respective results of QGSJET-III and other
  models are analyzed. In particular, it is demonstrated that those differences
  are partly caused by severe deficiencies of the other interaction models.
}

\vspace{10pt}
\noindent\rule{\textwidth}{1pt}
\tableofcontents\thispagestyle{fancy}
\noindent\rule{\textwidth}{1pt}
\vspace{10pt}

\section{Introduction}
\label{sec:intro}
Studies of cosmic rays (CRs) of very high energies are traditionally performed using the extensive air shower (EAS) techniques: measuring characteristics of
nuclear-electromagnetic cascades induced by interactions of the primary CRs in the atmosphere. Consequently, a necessary ingredient of the
corresponding experimental analysis procedures  are numerical simulations of
EAS development. A special role in such simulations is played by Monte Carlo (MC)
generators of high energy hadronic interactions, designed to treat inelastic
collisions with air nuclei of both the primary CR particles and of secondary
hadrons produced in the course of EAS development.

Since general collisions of hadrons and nuclei can not be fully described within
the perturbative framework, such MC generators necessarily involve phenomenological
approaches. Therefore, the role of the calibration of CR interaction models, based
on available accelerator data, notably, from the Large Hadron Collider (LHC),
is difficult to overestimate. On the other hand, of equal importance is an
overall self-consistency of the underlying theoretical approaches employed in such
models. Needless to say, respecting the relevant conservation laws, e.g.\ 
regarding   energy-momentum and electric charge, and symmetries (e.g.\ 
isospin\footnote{While the isospin symmetry is not an exact one for strong
interactions, it holds to a very good accuracy thanks to the small mass difference
between the $u$ and $d$ quarks.}) is a must. Doing otherwise leads one to incorrect
predictions for various relevant characteristics of hadronic interactions and may
introduce an important bias into CR data analyses.


\section{Quark-Gluon String and Dual Parton models}
\label{sec:qgs}
All present MC generators of hadronic interactions are based on the qualitative
picture of quantum chromodynamics (QCD): the collisions between both hadrons and
nuclei are mediated by cascades of partons [(anti)quarks and gluons]. It is 
important to keep in mind that interacting hadrons form their parton ``coats'' 
well before the collision: by emitting multiple virtual parton cascades.
When some partons from the projectile  ``cloud'' meet their counterparts from the
target and scatter of each other, the scattering may destroy the coherence of the
initial parton fluctuation, causing inelastic rescattering processes giving rise
to secondary hadron production. Alternatively, the coherence of some virtual
parton cascades may be preserved by the scattering process and the corresponding
partons will recombine back to their parent hadrons, which corresponds to
elastic rescattering processes. Thus, a hadron-hadron interaction generally
contains   multiple inelastic   and elastic rescattering processes.

It is customary to use the eikonal approximation for treating multiple
scattering: assuming all the inelastic and elastic processes to be independent
of each other, for a given impact parameter $b$ between the interacting hadrons.
This leads one  to simple expressions for the interaction cross sections,
e.g.\ for the total proton-proton cross section one obtains
\begin{equation}
\sigma^{\rm tot}_{pp}(s) = 2\int \! d^2b \left[1-e^{-\chi_{pp}(s,b)}\right],
\label{eq:sigtot}
\end{equation}
where $s$ is the center-of-mass (c.m.) energy squared for the collision and
the eikonal $\chi_{pp}(s,b)$ is defined by the imaginary part of the scattering
amplitude for a single 
 scattering process.

A very successful description of high energy hadronic collisions had been 
provided by the Quark-Gluon String   \cite{kai82} and Dual Parton \cite{cap91}
models developed within the Reggeon Field Theory (RFT) framework \cite{gri68}.
An elementary rescattering process has been described by a Pomeron exchange,
with  the respective eikonal $\chi_{pp}^{\mathbb P}$ having only  3 adjustable parameters:
\begin{equation}
\chi_{pp}^{\mathbb P}(s,b)=\frac{\gamma_p^2\,
s^{\Delta}}{2R_p^2+\alpha_{\mathbb P}'\,\ln s}\,
\exp \!\left[- \frac{b^2/4}{2R_p^2+\alpha_{\mathbb P}'\,\ln s}\right].
\label{eq:chi-pom}
\end{equation}
 The so-called
overcriticality $\Delta >0$ controls the energy-rise of   $\chi_{pp}^{\mathbb P}$, reflecting the
energy-rise of the parton density, while the Pomeron slope $\alpha_{\mathbb P}'$
is related to parton transverse diffusion.

The conversion of partons into secondary hadrons involved the concept of the
color-exchange: after the collision, constituent partons [(anti)quarks and
(anti)diquarks] from the interacting hadrons appeared to be connected to each
other by tubes (strings) of color field.
 With the partons flying apart, the tension of the string rises until it breaks,
with the color field being neutralized via a creation of additional 
quark-antiquark and diquark-antidiquark pairs from the vacuum, giving rise to
a formation of  secondary hadrons.

\section{From QGSJET to QGSJET-III}
\subsection{Combined treatment of soft and hard processes}
\label{sec:qgs1}
The traditional RFT assumes hadronic collisions to be dominated by pure soft
processes, corresponding to production of hadrons of relatively low
transverse momenta $p_{\perp}\lesssim 1$ GeV. On the other hand, with increasing
energy, the so-called semihard processes involving cascades of high  $p_{\perp}$
partons become more and more important \cite{glr}. This is because the smallness
of the respective strong coupling $\alpha_{\rm s}(p_{\perp}^2)$ becomes 
compensated by large collinear and infrared logarithms: the logarithmic ratios
of transverse $\ln (p_{\perp _{i}}^2/p_{\perp _{i-1}}^2)$ and longitudinal
$\ln (x_{i-1}/x_i)$ momenta of subsequent partons [$x_i$ being the fraction
of the parent hadron light cone (LC) momentum, carried by $i$-th parton]
 in the corresponding parton
cascades preceding the hardest (highest $p_{\perp}$) parton-parton scattering.

The QGSJET model \cite{kal94,kal97} was designed to treat both soft and semihard
processes coherently within the RFT framework, based on the ``semihard Pomeron''
approach  \cite{kal94,dre99,dre01}.
The main idea was to employ the 
perturbative QCD (pQCD) formalism for treating perturbative parts of the 
underlying parton cascades, for parton virtualities $|q^2|$ above some chosen
cutoff $Q_0^2$ for pQCD being applicable, while keeping the Pomeron description
for pure soft ($|q^2|<Q_0^2$) processes and for soft parts of semihard parton
cascades.
This allowed one to develop the Pomeron calculus, based on the ``general
Pomeron'' which thus combines the soft and semihard contributions.

With regard to EAS modeling, the main consequence of
taking semihard processes into account was a steeper energy rise of the 
multiple scattering rate: $\propto s^{\Delta_{\rm hard}}$,
 $\Delta_{\rm hard}\simeq 0.3$, compared to the one for pure soft processes
 ($\propto s^{\Delta_{\rm soft}}$,  $\Delta_{\rm soft}\simeq 0.1$); the patterns
 of secondary hadron production in the projectile fragmentation region
 dominating the EAS development do not differ significantly between soft and
 semihard inelastic rescatterings.
 
 The latter point deserves an additional discussion. Naively, one may question
 the importance of relatively high $p_{\perp}$ jet production for EAS modeling: 
 since such jets are typically produced in the central $y\sim 0$ rapidity region in
 c.m.\ frame, having therefore a minor influence on forward hadron production.
 However, the crucial role here is played by the parton cascades preceding the
 hardest  parton-parton scattering: each previous parton
 in such a cascade is characterized by a smaller transverse momentum, compared
 to the subsequent one, $p_{\perp _{i-1}}\ll p_{\perp _{i}}$, and a higher LC
 momentum fraction, $x_{i-1}\gg x_i$. Therefore, of highest importance 
 for EAS development are the
 partons produced in the very beginning of such cascades. In the semihard
 Pomeron approach, those cascades start already in the nonperturbative region
 ($p_{\perp}<Q_0$), hence, at large $x$, thereby having a strong impact on the
 respective EAS predictions, as discussed in some detail in Ref.\  \cite{ost16}.
 
 A striking counter-example is the approach of the SIBYLL model
  \cite{fle94,rie20}, which ignores the existence of such cascades and 
  takes into consideration the highest  $p_{\perp}$ parton-parton scattering
  only. Obviously, this is wrong from first principles: those are such parton
  cascades which produce the above-discussed collinear and infrared 
  enhancements of high   $p_{\perp}$ jet production, being therefore the very
  reason for the energy-rise of the jet production rate. On the other hand,
  from the pragmatic point of view, this leads to a serious underestimation
  of secondary hadron production in the fragmentation regions, giving rise to
  contradictions with LHC data and to incorrect predictions for EAS
  characteristics  \cite{ost16}.

\subsection{Microscopic treatment of nonlinear interaction effects}
\label{sec:qgs2}
The next crucial step was to 
consider nonlinear effects related to interactions between the ``elementary''
parton cascades: treating those as Pomeron-Pomeron interactions and performing
all-order resummations of the respective multi-Pomeron graphs
\cite{ost06,ost08,ost10}. This formed the basis for the development of the
QGSJET-II model \cite{ost06b,ost11,ost13}, allowing one both to calculate
various cross sections for high energy hadronic collisions and to perform
MC simulations of inelastic interaction events, generating the (generally
complicated) event topologies in strict correspondence with the respective
partial cross sections \cite{ost11}.

The corresponding microscopic treatment involved a single additional adjustable
parameter, the triple-Pomeron coupling, whose value was constrained based on
 HERA measurements of diffractive structure functions \cite{ost06b,ost11}. On the
 other hand, it produced a rich phenomenology characterized by numerous
 nontrivial dynamical effects regarding, e.g.\ a coherent description of proton
 structure functions and the energy-dependence of $\sigma^{\rm tot}_{pp}$
 \cite{ost06b}, stronger nonlinear
  effects in proton-nucleus and nucleus-nucleus collisions \cite{ost11},
 the energy-dependence of multi-parton scattering rates \cite{ost16c} and of the
 rapidity gap suppression  \cite{ost18}.

\subsection{Higher twist corrections to hard scattering processes}
\label{sec:qgs3}
A new theoretical mechanism implemented in the QGSJET-III model  \cite{ost19,ost21}
concerned a treatment of the so-called power corrections to hard parton-parton
scattering processes. In all present MC generators of hadronic collisions,
the description of hard processes is based on the leading twist pQCD
factorization \cite{col88}. In particular, the hardest
 process in that formalism corresponds to a binary parton-parton
scattering  involving a single parton  from the
projectile hadron (nucleus) and a single one from the target. While such a
formalism is fully justified  for high enough transverse momenta, it is
expected to break down at moderately small $p_{\perp}$, where higher twist
corrections become potentially important. Consequently, current MC generators
face a problem of an uncontrollable rise of the jet production rate in the 
small $p_{\perp}$ limit, which leads to a  strong sensitivity of
model results to the choice of the above-discussed $Q_0^2$-cutoff separating
the   treatments of hard and soft processes.

An important class of  higher twist corrections corresponding to coherent
rescattering of produced $s$-channel partons on ``soft'' (small LC momentum)
gluon pairs has been identified in Refs.\ \cite{qiu04,qiu06}. The respective
hard scattering processes thus involve arbitrary numbers of partons from the
projectile or target hadrons (nuclei). Since the corresponding multi-parton
correlators have not been measured experimentally, a model implementation
of the approach necessarily implies a phenomenological treatment.

\begin{figure}[h]
\centering
\includegraphics[width=0.40\textwidth]{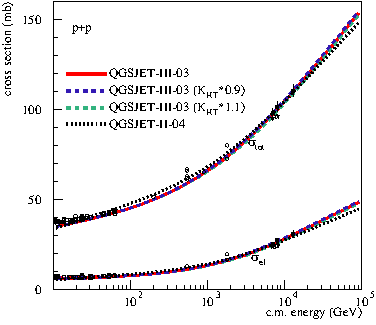}
\caption{$\sqrt{s}$-dependence of  the total and elastic $pp$ cross sections,
 calculated with the QGSJET-III model,
for  the default value of
$K_{\rm HT}$ and for $\pm 10$\% variations of that parameter, 
compared to QGSJET-II-04  results  \cite{ost11,ost13} and to experimental data.}
\label{fig:sig}
\end{figure}

In QGSJET-III, such  multi-parton correlators have been interpreted
probabilistically: as the so-called generalized multi-parton distributions,
which allowed one to develop a dynamical microscopic treatment of the 
corresponding nonlinear effects,  introducing a single additional
adjustable parameter $K_{\rm HT}$ which controls the magnitude of such higher
twist corrections \cite{ost19,ost19a}. Among the consequences of this new
mechanism is a drastic reduction of the model sensitivity to the choice of the
 $Q_0^2$-cutoff, taming the energy rise of  the interaction cross sections
 and of secondary hadron multiplicity, a stronger damping of low $p_{\perp}$
 jet production in more central collisions, etc.
Regarding the parameter $K_{\rm HT}$, the model results are not too sensitive
to its precise value, as illustrated in Fig.\ \ref{fig:sig}, 
 where the calculated $\sigma^{\rm tot}_{pp}$  and  $\sigma^{\rm el}_{pp}$
 are plotted both for the default value of
$K_{\rm HT}$ and for $\pm 10$\% variations of that parameter.

\subsection{Pion exchange process}
\label{sec:qgs3-pi}
An additional technical improvement in the QGSJET-III model concerned a
treatment of the pion exchange process in hadronic collisions \cite{ost21}.
The importance
of that process for calculations of the EAS muon content $N_{\mu}$ has been 
stated in Ref.\ \cite{ost13}: in pion-air collisions, the $t$-channel pion
exchange   enhances forward production of $\rho^0$ mesons,
by the expanse of neutral pions, which leads to a $\sim 20$\% increase of the
predicted muon density at ground level. This provides a sufficient motivation
to develop a consistent treatment of the mechanism and to cross check the
formalism against the data of the LHCf experiment, regarding forward neutron
production in $pp$ collisions.
The main theoretical challenge here is to predict
the energy-dependence of the process, which is governed by the corresponding
absorption effects: since those define the probability for not filling the
rapidity gap between the forward produced neutron in   $pp$ collisions or the
forward  $\rho^0$ in pion-proton (pion-nucleus) interactions and the other
secondary hadrons produced. While the details on the corresponding treatment
can be found elsewhere \cite{ost21}, it may be instructive to compare the
predicted energy-dependence of forward  $\rho^0$  production in pion-nitrogen
collisions, between QGSJET-III and   other CR interaction models.

\begin{figure}[h]
\centering
\includegraphics[width=0.95\textwidth]{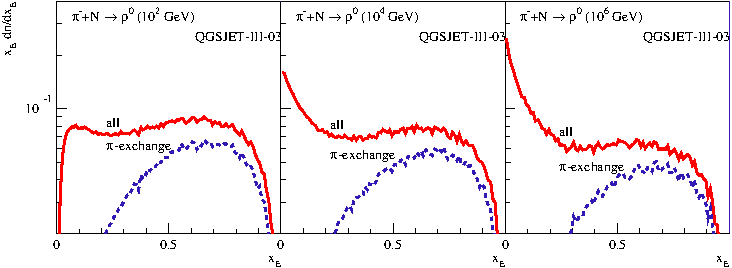}
\caption{Energy spectra of  $\rho^0$ mesons produced in $\pi^- -^{14}\!\!N$ interactions at
$10^2$ (left), $10^4$ (middle), and $10^6$ (right) GeV, calculated with the  QGSJET-III
model;  solid lines - total spectrum, dashed lines - 
contribution of the pion exchange process.}
\label{fig:rho0-qiii}
\end{figure}

\begin{figure}[h]
\centering
\includegraphics[width=0.81\textwidth]{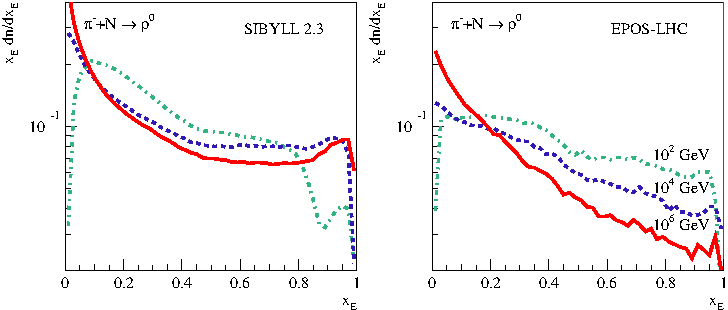}
\caption{Energy spectra of  $\rho^0$ mesons produced in $\pi^- -^{14}\!\!N$ interactions at
$10^2$ (dashed-dotted), $10^4$ (dashed), and $10^6$ (solid) GeV, calculated with  
the SIBYLL-2.3 (left) and  EPOS-LHC (right) models.}
\label{fig:rho0-sib}
\end{figure}

As one can see in Fig.\ \ref{fig:rho0-qiii}, 
the pion exchange process in QGSJET-III
dominates indeed the forward  $\rho^0$ yield and the corresponding contribution
slowly decreases with energy, being stronger and stronger damped by the
above-discussed absorption effects, which is a direct consequence of the
energy-rise of the multiple scattering rate. A similar but  stronger 
damping of the  forward  $\rho^0$ yield is predicted by the EPOS-LHC 
model \cite{pie15}, see  Fig.\ \ref{fig:rho0-sib} (right).
On the other hand,  in the case of  the SIBYLL-2.3 model \cite{rie20},
 no pronounced damping with increasing energy
is observed for the  forward  $\rho^0$ yield,
as is easy to see in Fig.\ \ref{fig:rho0-sib} (left).
This indicates that the relevant  absorption effects are seriously
underestimated, which may lead to an artificial enhancement of the predicted
EAS muon content.

\section{EAS predictions: QGSJET-III and other models}
\label{sec:EAS}

\begin{figure}[h]
\centering
\includegraphics[width=0.37\textwidth]{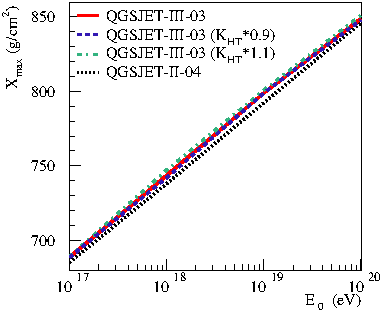}
\caption{Energy-dependence of the shower maximum depth for proton-induced EAS,
 calculated with the QGSJET-III model, for  the default value of
$K_{\rm HT}$ and for $\pm 10$\% variations of that parameter, compared to the results of
QGSJET-II-04.}
\label{fig:xmax}
\end{figure}

For basic EAS characteristics, rather small differences have been observed between the 
predictions of QGSJET-III and of the previous model version, QGSJET-II-04. For example,
for the average shower maximum depth $X_{\max}$, those amount to some 5 g/cm$^2$, as
one can see in  Fig.\ \ref{fig:xmax}.
Varying the $K_{\rm HT}$ parameter by $\pm 10$\% produces only  $\pm 2$ g/cm$^2$
changes of  $X_{\max}$. Even smaller differences ($\sim 1$\%) between  QGSJET-III 
and  QGSJET-II-04 have been observed for the EAS muon content $N_{\mu}$.

While the reasons for a potential stability of  the predicted  $N_{\mu}$ will be
discussed elsewhere, let us concentrate here on various model predictions for $X_{\max}$.
 In principle, a robustness of the respective results is expected if: (i) those are
 dominated by the treatment of proton-air collisions and (ii) such a treatment is
 sufficiently constrained by LHC data, notably, by measurements of the total and
 elastic $pp$ cross sections.
 
 Given the higher  $X_{\max}$ values predicted by the EPOS-LHC and SIBYLL-2.3 models,
 one may question the validity of both above assumptions. However, a slower EAS development
 predicted by SIBYLL is a direct consequence of the general theoretical pathology of that
 model, regarding the treatment of hard processes in hadronic collisions, as discussed in
 Section \ref{sec:qgs1}. More interesting is the case of the  EPOS-LHC model: while there seems to
 be no general problem with the model approach, a number of its technical features appear
 to be  questionable, e.g.\ an enhanced forward production of (anti)baryons in
 pion-proton and pion-nucleus collisions. 
 As demonstrated in Ref.\ \cite{ost16a},
 the latter has some impact on the predicted  $X_{\max}$: shifting the shower maximum depth
 somewhat deeper in the atmosphere. Since this questions the assumption (i) above,
 let us 
 have a closer look at the matter.
 
\begin{figure}[h]
\centering
\includegraphics[width=0.85\textwidth]{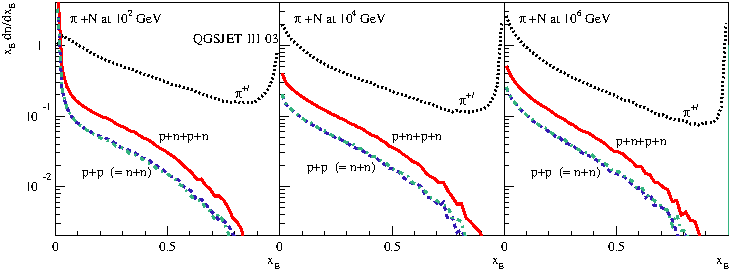}
\caption{Energy spectra of charged pions and (anti)baryons produced 
in $\pi^- N$ 
collisions at $10^2$ (left), $10^4$ (middle), and $10^6$ (right) GeV, 
calculated with    QGSJET-III.}
\label{fig:pbar-qiii}
\end{figure}

\begin{figure}[h]
\centering
\includegraphics[width=0.85\textwidth]{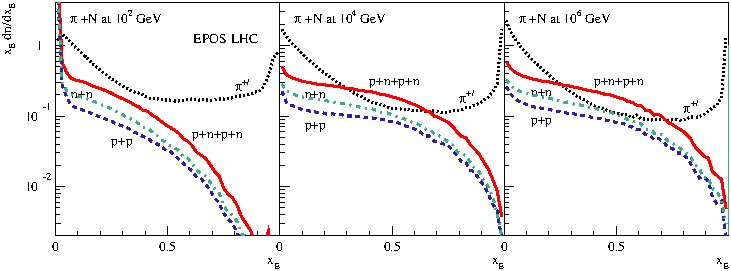}
\caption{Same as in Fig.\ \ref{fig:pbar-qiii}, calculated with the  EPOS-LHC model.}
\label{fig:pbar-epos}
\end{figure}

In Figs.\ \ref{fig:pbar-qiii} and \ref{fig:pbar-epos},
we compare the energy  spectra of charged pions and (anti)baryons in pion-nitrogen collisions at
$10^2$, $10^4$, and $10^6$ GeV, predicted by  QGSJET-III and  EPOS-LHC. While in the former model,
the forward  (anti)baryon yield at all the energies is suppressed by some order of magnitude,
compared to the one of pions, a strikingly different behavior is predicted by EPOS: the 
forward  (anti)baryon production steeply rises with energy. Neglecting the very forward
($x_E\rightarrow 1$) part of the pion spectra, which is dominated by   contributions of
diffractive processes, the forward  (anti)baryon yield in pion-nucleus collisions,
 predicted by EPOS-LHC, appears to exceed
significantly the one of charged pions, in the very high energy limit.
 While such a picture can not be excluded by general arguments, it seems to be an artificial one:
 since there exists no viable theoretical mechanism to produce such an effect.
 
An additional
 enhancement of the  forward  (anti)baryon yield in  EPOS-LHC
 arises from non-respecting the isospin symmetry by that model: as one can see in 
 Fig.\  \ref{fig:pbar-epos}, its forward $(n+\bar n)$ spectra 
exceed substantially
 the ones of protons plus antiprotons (c.f., Fig.\ \ref{fig:pbar-qiii} for the respective
 results of QGSJET-III). Obviously, 
 this is wrong from first principles.

\section{Relevance to UHECR composition studies}
\label{sec:pao}
At this point, it is worth to discuss the consequences for
 experimental studies of the composition of ultra-high energy cosmic rays
  (UHECRs), 
arising from a potential robustness of the model predictions for EAS characteristics.
It may be instructive to consider the respective results of the Pierre Auger Observatory (PAO),
resulting from measurements  of the EAS maximum depth  $X_{\max}$ \cite{pao14}. 
A peculiar feature of the corresponding  data is a drastic decrease of $X_{\max}$ fluctuations,
$\sigma(X_{\max})$, 
in the ultra-high energy limit \cite{pao14a}.
Consequently, to interpret coherently the experimental measurements of both the average
$X_{\max}$  and  $\sigma(X_{\max})$, one favors CR interaction models which predict a deeper
shower maximum depth or/and smaller fluctuations of $X_{\max}$. In particular, the best
 overall consistency had been stated for  EPOS-LHC, mostly because of the smaller
 $\sigma(X_{\max})$   predicted by that model for nucleus-induced EAS.
 
 In reality,  $\sigma(X_{\max})$ is a theoretically robust quantity, as stated already in 
 Ref.\ \cite{alo08}.
 The small fluctuations of the shower maximum depth,
 predicted by  EPOS-LHC for nucleus-initiated air showers, result from an erroneous
 treatment of nuclear break up by that model\footnote{Regarding the importance
 of the process for the fluctuations of $X_{\max}$, see Ref.\ \cite{kal93}.} \cite{ost16b,ost19c}.
In turn, as discussed  in Section \ref{sec:EAS}, the higher $X_{\max}$ values predicted by
SIBYLL-2.3 and EPOS-LHC are, at least partly, caused by deficiencies of those models.
This rises the question on how the PAO results on UHECR composition would change if the
experimental data were reinterpreted using corrected or/and alternative CR interaction models.

\section{Outlook}
\label{sec:Outlook}
We discussed  the main theoretical approaches implemented in the
QGSJET-III MC generator, comparing also to alternative
treatments of the other CR interaction models and revealing a number of
serious deficiencies of the latter. As for QGSJET-III, while its predictions
are largely driven by the underlying theoretical mechanisms, the model
necessarily involves various phenomenological assumptions, 
which makes its results unwarranted.

Regarding the predictions of  QGSJET-III for basic EAS characteristics, those
appeared to differ little from the ones of the previous model version, 
 QGSJET-II-04, which may indicate that such predictions are already constrained
 substantially by available accelerator data, notably, from LHC. In relation to that,
 we demonstrated that the different EAS predictions of the other
  CR interaction models can, at least partly, be explained
   by  deficiencies of those models. 
  Therefore,  real
  uncertainties for  EAS predictions are very likely smaller than the
  differences between the  results of the present models.
  A  more definite statement   requires a thorough quantitative
  study of such uncertainties, which appears to be  an urgent and important  task.

%


\paragraph{Funding information}
This work was supported by  Deutsche
 Forschungsgemeinschaft (project number 465275045).

\bibliography{ostapchenko_isvhecri2022.bib}

\nolinenumbers

\end{document}